\documentclass[twocolumn]{aastex701}

\usepackage{amsmath,amssymb,bm,physics}
\usepackage{graphicx,booktabs,multirow}
\usepackage{xcolor}

\newcommand{\gadget}{\textsc{Gadget-4}}

\begin{document}

\title{TUNeS: Neural Emulation of Large-Scale Structure Across Redshifts}

\author[orcid=0009-0007-2932-0437]{Yuqi Kang}
\affiliation{Institute for Frontier in Astronomy and Astrophysics, Beijing Normal University, Beijing, 102206, China}
\affiliation{School of Physics and Astronomy, Beijing Normal University, Beijing 100875, China}
\email[]{yuqi.kang@bnu.edu.cn}  

\author[orcid=0000-0001-5093-8118]{Bin Hu} 
\affiliation{Institute for Frontier in Astronomy and Astrophysics, Beijing Normal University, Beijing, 102206, China}
\affiliation{School of Physics and Astronomy, Beijing Normal University, Beijing 100875, China}
\email[show]{bhu@bnu.edu.cn}

\author[orcid=0000-0001-9653-0337]{Dongxing Li} 
\affiliation{Faculty of Geographical Science, Beijing Normal University, Beijing 100875, China}
\email[]{lidx@bnu.edu.cn}

\author[orcid=0000-0003-2006-0089]{Jan Hamann}
\affiliation{Sydney Consortium for Particle Physics and Cosmology, School of Physics, The University of New South Wales, Sydney NSW 2052, Australia}
\email[]{jan.hamann@unsw.edu.au}

\begin{abstract}
In this work, we introduce TUNeS (Temporal UNet emulator for Structure formation), a neural network framework for accelerating N-body simulations by predicting the nonlinear evolution of the matter density field from an initial particle distribution. TUNeS employs a two-stage modeling strategy, combining particle-based inference with a density-field refinement on a regular grid, enabling accurate reconstruction of both large- and small-scale structures. The model is designed to operate across redshift, taking particle snapshots at arbitrary input redshifts and predicting density fields at arbitrary target redshifts. In this work, we evaluate its performance using simulations initialized at $z=100$, with predictions generated at multiple lower redshifts. Trained on only eight N-body simulations, TUNeS reproduces reference results with good agreement in both Gaussian and non-Gaussian statistics, including two-point correlations, one-point distributions, peak counts, and three-dimensional Minkowski functionals. In particular, at $k \simeq 1\,h\,\mathrm{Mpc}^{-1}$, the power spectrum error remains at the few-percent level. End-to-end inference from $256^3$ particles to a $256^3$ density grid can be completed in $\sim25\,\mathrm{second}$ on a single GPU. 
Thanks to its architectural design, the model naturally scales to larger particle numbers and larger volumes through particle batching and window-based refinement.
\end{abstract}

\section{Introduction}\label{sec:intro}

In recent years, an increasing number of ongoing and forthcoming cosmological surveys have begun to deliver high-precision observational data over unprecedented volumes and redshift ranges ~\citep{LSSTScience:2009jmu,EUCLID:2011zbd,DESI:2016fyo,SimonsObservatory:2018koc,CSST:2025ssq}. The full exploitation of these data requires large ensembles of numerical simulations, which are essential to making theoretical predictions, estimating covariance matrices, and characterizing systematic uncertainties ~\citep{Hartlap:2006kj,Dodelson:2013uaa,Petri:2016wlu}.

Among the available approaches, N-body simulations constitute the standard and most reliable tool for modeling the nonlinear gravitational evolution of large-scale structures~\citep{Springel:2000yr,Springel:2017tpz,Springel:2020plp}. By directly solving the equations of motion for a large number of particles under gravity, N-body simulations accurately capture nonlinear effects that are inaccessible to perturbative methods~\citep{Bernardeau:2001qr}. However, this accuracy comes at a substantial computational cost. High-resolution simulations, which are crucial for resolving small-scale nonlinear structures, demand both a large number of particles and fine spatial and temporal resolution, leading to prohibitive requirements on computational time and data storage. Therefore, in practice, N-body simulations are often constrained by available computational resources, forcing a trade-off between simulation volume, mass resolution, and the number of realizations.

To mitigate the high computational cost of full N-body simulations, a variety of approximate and accelerated methods have been developed over the past decades. Representative examples include particle-mesh (PM) methods ~\citep{1988csup.book.....H}, adaptive mesh refinement (AMR) methods~\citep{2002A&A...385..337T,2014ApJS..211...19B}, approximations based on Lagrangian perturbation theory ~\citep{1994A&A...288..349B,1998MNRAS.299.1097S,Bernardeau:2001qr}, and hybrid schemes such as COLA~\citep{Tassev:2013pn}, FastPM~\citep{Feng:2016yqz} or BullFrog~\citep{Rampf:2024uvj}. By simplifying gravitational force calculations and time integration, these approaches significantly reduce computational expense. However, their precision is ultimately constrained by the underlying approximations, particularly in the deeply nonlinear regime and on small scales~\citep{Schneider:2015yka,Bayer:2025ija}. 

In parallel with these developments, the rapid advancement of machine learning (ML) techniques has opened new avenues for accelerating numerical simulations in a wide range of scientific disciplines, including chemistry\citep{Vamathevan2019}, biomedical imaging ~\citep{Litjens2017} and physics~\citep{Albertsson:2018maf,Carleo:2019ptp}. In the context of cosmological large-scale structure formation, neural networks have emerged as a powerful tool for emulating N-body simulations. By learning the complex and highly nonlinear mapping between initial conditions and evolved matter distributions from training data, these models can produce accurate predictions at a fraction of the computational cost and with orders of magnitude shorter run times compared to traditional N-body simulations.Pioneering studies demonstrated that convolutional neural networks (CNNs) can learn the nonlinear mapping from initial conditions to late-time matter distributions~\citep{He:2018ggn}. 
Subsequent works have explored a wide range of approaches, including particle-based neural networks that map linear displacement fields to nonlinear counterparts~\citep{AlvesdeOliveira:2020yix,Jamieson:2022lqc,Jamieson:2024fsp}, fully differentiable cosmological simulation frameworks using the adjoint method~\citep{Li:2022bsu}, generative super-resolution models~\citep{Li:2020vor,Ramanah:2020vyl,Hafezianzadeh:2025ifw}, and diffusion frameworks~\citep{Rouhiainen:2023ewv,Riveros:2025aec,Mishra:2026xsl} that synthesize nonlinear density fields with high fidelity. Together, these efforts demonstrate the growing potential of machine learning as a scalable and flexible tool to accelerate simulations of the formation of cosmological structures.

In this work, we introduce TUNeS (Temporal UNet emulator for Structure formation)\footnote{The code and trained models for TUNeS are publicly available at \url{https://github.com/Kang-Yuqi/TUNeS}.}, a neural-network-based framework for accelerating simulations of cosmological structure formation that operates directly on particle data. The proposed model takes particle positions and velocities at an initial redshift as input, avoiding dependence on grid-based initial conditions or linear perturbation–based displacement schemes. Instead of predicting particle trajectories explicitly at all times, the model maps the initial particle configuration to the evolved three-dimensional matter density field at a target redshift.

Our approach consists of two stages. In the first stage, a particle-based network provides a coarse prediction of particle displacements, capturing the linear-like evolution of large-scale structures in a computationally efficient manner. In the second stage, the predicted particle distribution is converted into a spatial density field, which is further refined using a three-dimensional U-Net architecture~\citep{Ronneberger:2015cwd} to model nonlinear evolution and small-scale structure formation. This hybrid design balances the physical interpretability of particle-based modeling with the strong representational ability of convolutional neural networks to capture highly clustered density fields.

The model is trained to produce redshift-dependent outputs over a wide range of high to low redshift. Unlike grid-based approaches that rely on regular lattice initial conditions, our framework places no restriction on the spatial configuration of the input particles and can operate on general particle distributions. The starting redshift is therefore not fixed by the model architecture itself, allowing simulations to be initialized at different cosmic epochs. Unlike conventional approaches that require evolving from extremely high redshifts and thus inherit assumptions about early-time physics, this flexibility enables the model, in principle, to propagate density fields forward from arbitrary intermediate redshifts. In this work, however, all results are obtained using a fixed input redshift of $z = 100$, which serves as a controlled and representative reference configuration for demonstrating the performance of the model.

To further reduce computational cost and enhance scalability, we introduce a windowing-based strategy that allows the trained model to be applied to arbitrarily large volumes without requiring periodic boundary conditions.

We validate the trained model using a comprehensive set of evaluation metrics that probe both Gaussian and non-Gaussian properties of the predicted matter density fields. These diagnostics include two-point statistics, one-point probability distribution functions, peak-based statistics, and morphological measures based on Minkowski functionals. Together, these tests are designed to assess the suitability of the model for precision large-scale structure applications, including weak gravitational lensing analyses and cosmological parameter inference.

This paper is organized as follows. In Section~\ref{sec:method}, we describe the overall methodology of the proposed framework, including the model overview, data representations, network architecture, loss functions, and training details. Section~\ref{sec:data} introduces the N-body simulations used in this work and outlines the data processing and reconstruction procedures, including density field window construction and window stitching. In Section~\ref{sec:result}, we present the main results, including visual comparisons and a series of statistical evaluations based on one-point, two-point, peak, and three-dimensional Minkowski functional statistics, as well as the runtime and memory performance of the model. Finally, we summarize our findings and discuss future directions in Section~\ref{sec:conclusion}.

\section{Method}\label{sec:method}

\begin{figure*}
    \centering
    \includegraphics[width=0.9\textwidth]{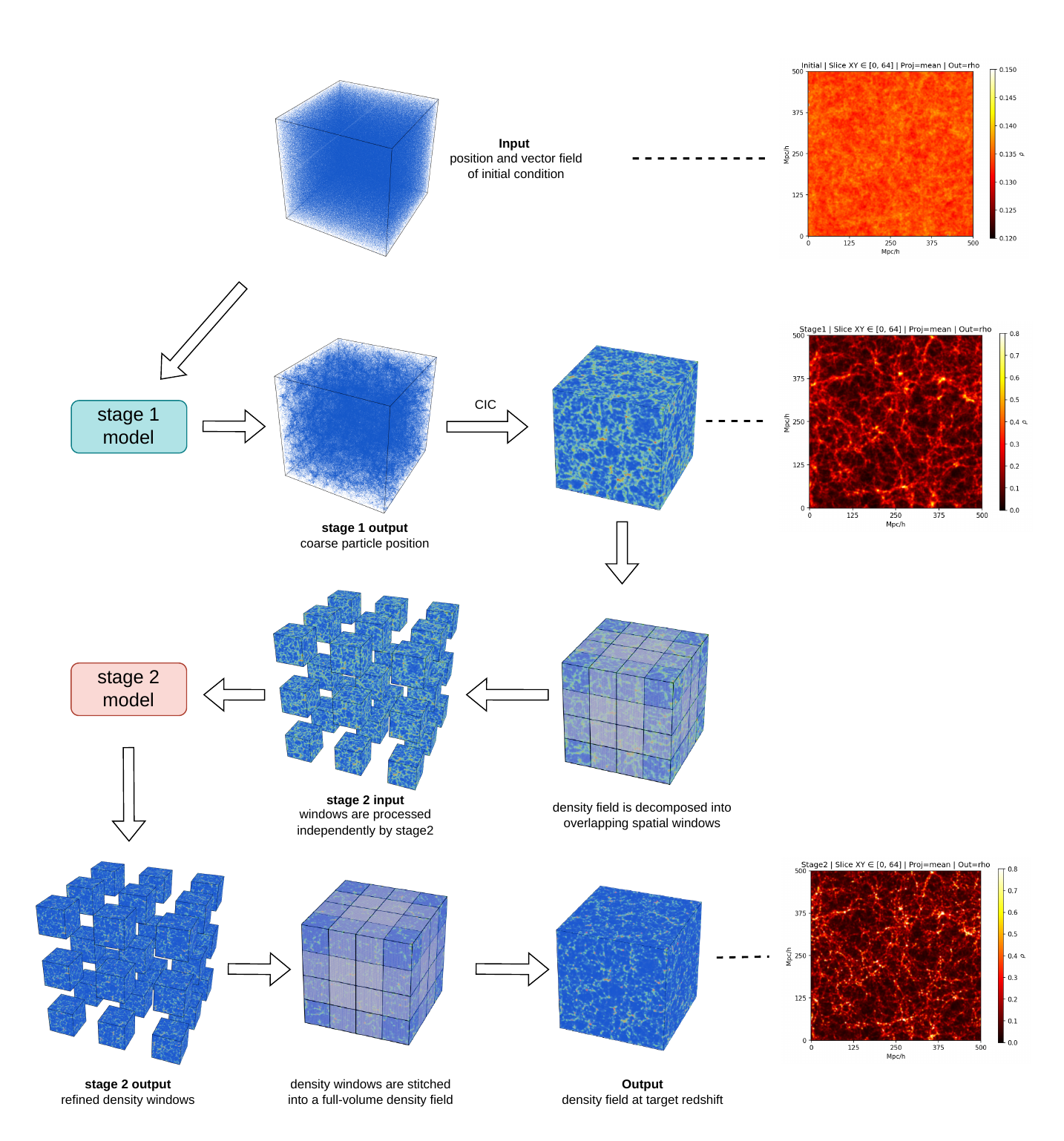}
    \caption{Overview of the pipeline for the proposed framework. The rightmost panels show two-dimensional mean density projections of slices with thickness $64\,h^{-1}\mathrm{Mpc}$ at different stages. For visualization, the particle field is randomly down-sampled to $10^6$ particles.}
    \label{fig:pipeline}
\end{figure*}

\subsection{Model Overview}\label{sec:overview}

The proposed framework adopts a two-stage modeling strategy that separates the treatment of large-scale, approximately linear evolution from the refinement of nonlinear, small-scale structures. This separation enables different stages to employ distinct neural-network architectures, with the first stage focusing on large-scale, long-wavelength modes associated with approximately linear particle displacements, and the second stage targeting small-scale nonlinear structures that are more effectively captured in image-based density representations. An overview of the proposed pipeline is illustrated in Fig.~\ref{fig:pipeline}.

In Stage~1, a particle-based model operates directly on the initial particle data to predict the particle displacements. This stage captures bulk motions and the overall structure formation driven by long-wavelength modes, providing a coarse yet physically interpretable approximation to the evolved matter distribution.

Stage~2 focuses on refining the small-scale and nonlinear features of the matter distribution. Operating on the density-field representation obtained from Stage~1, this stage employs a 3D U-Net convolutional neural network to model localized nonlinear evolution. By processing the density field in overlapping spatial windows and recombining the outputs, the model achieves accurate reconstruction of small-scale structures while maintaining scalability to large volumes.

\subsection{Data Representations}\label{sec:targets}

The proposed framework employs multiple data representations that correspond to different stages of the modeling process. The initial input consists of particle positions and velocities at a given starting redshift. In the first stage, the data are represented by the particles' phase-space coordinates, and a predicted displacement field is applied to the initial particle positions to obtain a coarse estimate of the particle distribution at the target redshift, with periodic boundary conditions applied when required.

The resulting particle distribution is subsequently mapped to a three-dimensional density field defined on a regular Cartesian grid. To enable localized modeling of small-scale nonlinear structures, this density field is further represented as a collection of overlapping spatial windows. These windowed density fields form the input to the second stage, where the model predicts refined density representations that encode small-scale nonlinear information. The window-level predictions are processed independently and later recombined to recover a continuous prediction over the full simulation volume. 

The detailed dataset construction procedures and subsequent data processing method will be described in Section~\ref{sec:data}.

\subsection{Network Architecture}\label{sec:arch}

\begin{figure*}
    \centering
    \includegraphics[width=\textwidth]{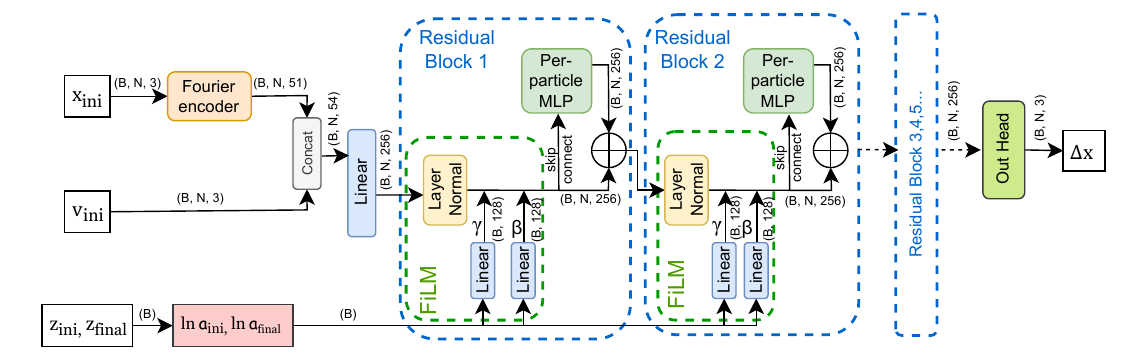}
    \caption{Architecture of the Stage-1 network. The model takes as input the initial particle positions $\boldsymbol{x}_{\rm ini}$, velocities $\boldsymbol{v}_{\rm ini}$, and the initial and target redshifts $(z_{\rm ini}, z_{\rm fin})$. Particle positions are encoded using Fourier features and concatenated with velocities, followed by a linear projection to a fixed-width hidden representation. Temporal information is embedded from the logarithmic scale factors and injected into each residual block via feature-wise linear modulation. The residual block, highlighted by a dashed blue frame, is repeated six times and consists of a FiLM-modulated layer normalization, a per-particle multilayer perceptron, and a residual skip connection. Feature dimensions at each stage are annotated along the arrows in the figure, and all residual blocks operate at the same feature dimension. The network outputs a per-particle displacement field $\Delta\boldsymbol{x}$. Here $B$ denotes the batch size and $N$ the number of particles.}
    \label{fig:stage1_model}
\end{figure*}

\paragraph{Stage~1}
As shown in Fig.~\ref{fig:stage1_model}, the Stage-1 model employs a feature-wise linear modulation (FiLM)~\citep{Perez2018FiLM} residual multilayer perceptron (MLP) architecture to predict coarse particle displacements. The model is designed to capture approximately linear gravitational evolution without having to explicit compute particle-particle interactions. 

For each particle, the input consists of its initial position $\boldsymbol{x}_{\rm ini}$, initial velocity $\boldsymbol{v}_{\rm ini}$, and the redshift pair $(z_{\rm ini}, z_{\rm fin})$. The initial positions are encoded using Fourier features to enhance the representation of large-scale spatial modes. The resulting positional embedding is concatenated with the particle velocity and linearly projected to a fixed-width hidden representation. 

Temporal information is embedded using the logarithmic scale factors $(\ln a_{\rm ini}, \ln a_{\rm fin})$ to form a global conditioning vector, which modulates the network through FiLM. Specifically, in each residual block the normalized feature $\boldsymbol{h}$ is modulated as
\begin{equation}
\mathrm{FiLM}(\boldsymbol{h}; \gamma, \beta) = (1 + \gamma)\odot \mathrm{LN}(\boldsymbol{h}) + \beta ,
\end{equation}
where the scale and shift parameters $(\gamma, \beta)$ are predicted from the time embedding through linear projections. Here, $\gamma$ and $\beta$ are vectors with the same dimensionality as the hidden feature $\boldsymbol{h}$. $\odot$ denotes the multiplication by element and $\mathrm{LN}$ represents the normal layer.

The FiLM-modulated features are then processed by a per-particle MLP with a residual skip connection. After a fixed number of such residual blocks (six in this work), a final prediction layer maps the hidden features to the per-particle displacement $\Delta \boldsymbol{x}$.

The Stage-1 prediction of particle positions is then obtained as 
\begin{equation} 
\boldsymbol{x}_{\rm stage1} = \boldsymbol{x}_{\rm ini} + \Delta \boldsymbol{x}, 
\end{equation} 
with periodic boundary conditions applied when necessary.

The depth of the residual MLP was determined empirically. We experimented with architectures ranging from three to seven residual blocks while keeping the hidden width fixed. We find that the training speed per epoch is largely insensitive to the depth within this range. Very shallow models ($<4$ blocks) show slightly degraded performance, indicating insufficient expressive capacity. We therefore adopt six residual blocks as a stable and computationally efficient choice.

Conceptually, the role of Stage~1 is related to the Zel’dovich approximation (ZA), which describes the linear gravitational evolution of particle displacements. In ZA, particle displacements are obtained by rescaling the initial displacement field with the linear growth factor. Similarly, the Stage~1 network aims to capture large-scale particle motions associated with approximately linear evolution. However, rather than assuming a fixed analytic form as in ZA, the Stage~1 network learns a direct mapping from the initial particle phase-space coordinates and the redshift pair to the particle displacement. This learned representation provides greater flexibility and a more general approximation to the large-scale dynamics, and can be naturally extended to alternative cosmological research such as modified gravity or multi-component matter models.

\begin{figure*}
    \centering
    \includegraphics[width=\textwidth]{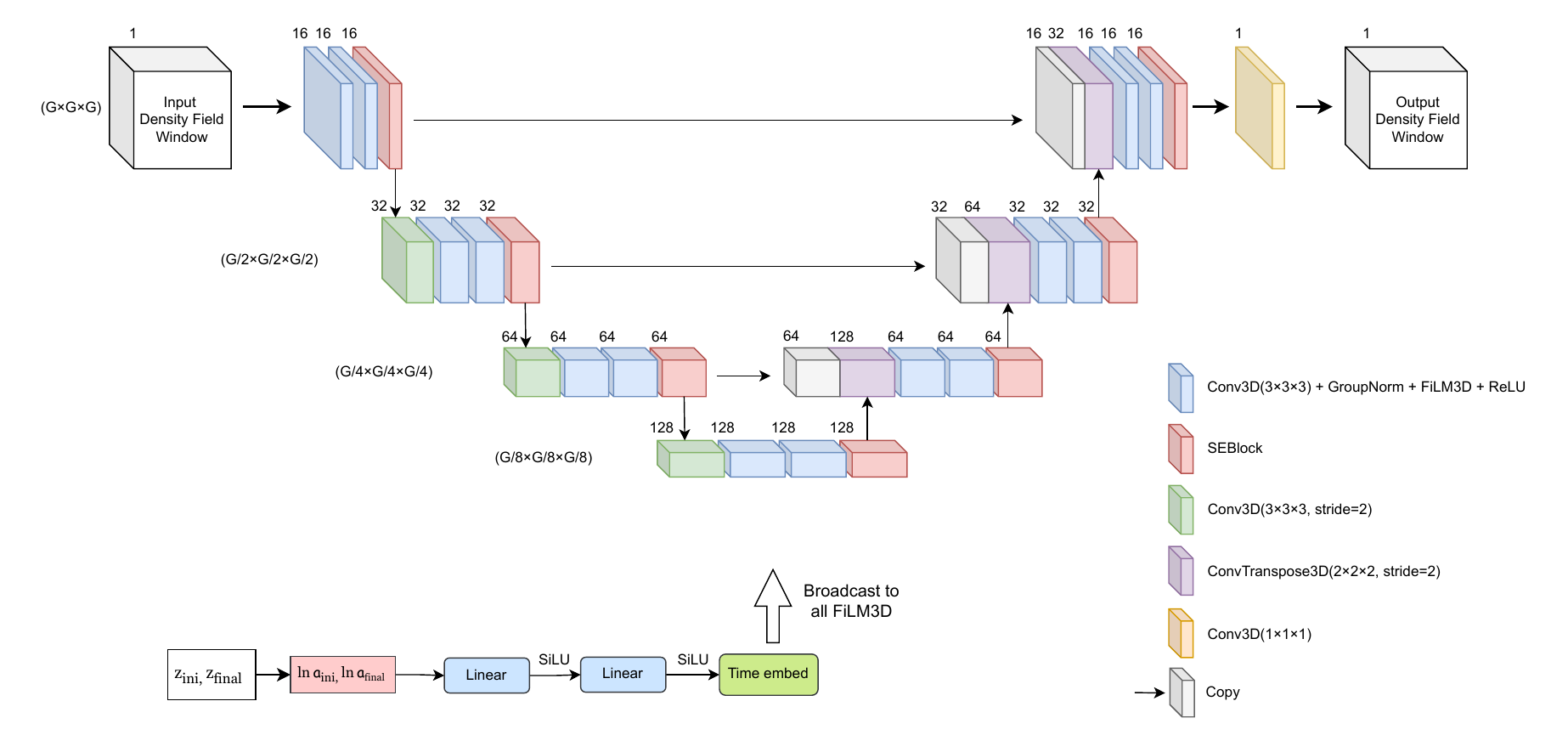}
    \caption{Stage-2 network architecture for density-field refinement. Colored blocks indicate different computational modules as defined in the legend. Strided and transposed 3D convolutions are used to change spatial resolution, while gray arrows denote skip connections that transfer feature maps across scales. Redshift information is embedded from $(\ln a_{\rm ini}, \ln a_{\rm fin})$ and broadcast to all FiLM3D layers, enabling redshift-dependent modulation throughout the network. The final output is obtained via a $1\times1\times1$ convolution producing a single-channel density-field window. The spatial resolution of each level is indicated on the left, and the label on top of each block denotes its number of channels.}
    \label{fig:stage2_model}
\end{figure*}

\paragraph{Stage~2}
As illustrated in Fig.~\ref{fig:stage2_model}, the Stage~2 model using a U-Net architecture~\citep{Ronneberger:2015cwd} to refine the coarse density field derived from the Stage~1 prediction, with redshift dependence explicitly encoded through 3D FiLM-based conditioning. The input to this stage are windowed three-dimensional coarse density fields constructed via cloud-in-cell (CIC) assignment from the Stage~1 prediction. The network outputs refined density fields that capture nonlinear evolution and small-scale structure.

The U-Net architecture follows a multi-level encoder-decoder design. In the encoder, the network progressively reduces the input density window using $3\times3\times3$ convolutions with stride 2 until it reaches the bottleneck. In the decoder, spatial resolution is restored by transposed $2\times2\times2$ convolutions for upsampling. At each resolution level, feature maps are transferred through skip connections implemented by channel-wise concatenation, which helps preserve spatial information and avoid vanishing gradients.

The spatial resolution at each level is constant, as shown in Fig.~\ref{fig:stage2_model}. At every resolution level, a convolutional block (ConvBlock3D) is applied to preserve spatial resolution while transforming the feature representation. For an input feature map $\boldsymbol{X}$, a ConvBlock3D can be expressed as
\begin{equation}
\boldsymbol{X}' = \mathrm{SE}\!\left( \mathrm{ReLU}\!\left( \mathrm{FiLM}\!\left( \mathrm{GN}\!\left( \mathrm{Conv}(\boldsymbol{X}) \right) \right) \right) \right),
\end{equation}
where $\mathrm{Conv}$ denotes a $3\times3\times3$ convolution, $\mathrm{GN}$ represents group normalization, $\mathrm{ReLU}$ is the activation function and $\mathrm{SE}$ denotes the squeeze-and-excitation operation~\citep{Hu2018SE}, which introduces adaptive channel-wise reweighting to enhance the representation of nonlinear features.

The redshift dependence is modeled through a three-dimensional feature-wise linear modulation (FiLM3D). Similarly to Stage~1, the initial and target redshifts are transformed into logarithmic scale factors $(\ln a_{\rm ini}, \ln a_{\rm fin})$ which are embedded in a global conditioning vector using a small multilayer perceptron composed of two linear layers with SiLU activations. For a feature map $\boldsymbol{X}$, FiLM modulation is applied as
\begin{equation}
\mathrm{FiLM}(\boldsymbol{X}; \gamma, \beta) = (1+\gamma)\odot \boldsymbol{X} + \beta,
\end{equation}
where the scale and shift parameters $(\gamma,\beta)$ are then broadcast to all ConvBlock3D layers.

The decoder output is projected to a single-channel density-field window using a $1\times1\times1$ convolution, which acts as a point wise linear mapping without introducing additional spatial mixing. The resulting output is the refined density field at the target redshift.

We tested architectures with three to five encoder–decoder levels while keeping the channel scaling pattern fixed. Networks with fewer than four levels exhibit slower convergence, whereas increasing the depth beyond four levels does not yield statistically significant improvements in validation metrics but incurs additional memory cost. We therefore adopt a four-level U-Net as a balanced configuration between accuracy and computational efficiency.

\subsection{Loss Functions}\label{sec:loss}

The model uses different loss functions at different stages due to the change in data type, from particle displacements in Stage~1 to windowed density-field pixel values in Stage~2.

\paragraph{Stage~1 Loss}
For the coarse particle displacement prediction in Stage~1, we employ a mean-squared-error (MSE) loss defined on the particle displacements,
\begin{equation}
\mathcal{L}_{\mathrm{stage1}} = \left\langle \left| \Delta\boldsymbol{x}_{\mathrm{s1}} - \Delta\boldsymbol{x}_{\mathrm{ref}} \right|^2 \right\rangle ,
\end{equation}
where $\Delta\boldsymbol{x}_{\mathrm{s1}}$ denotes the displacement predicted by the Stage~1 model, and $\Delta\boldsymbol{x}_{\mathrm{ref}}$ is the reference displacement obtained from the $N$-body simulation. The angle brackets indicate an average over all particles in the batch.

\paragraph{Stage~2 Loss}
In Stage~2, the input data are the density field windows defined on a Cartesian grid. Each window is trained independently, without enforcing explicit continuity across neighboring windows. To suppress artifacts introduced by window-based training and truncation at patch boundaries, we adopt a spatially weighted voxel-wise loss function.

Specifically, the loss gives a higher weight to central region voxels and down-weights voxels approaching the window boundaries. With an MSE base loss, the Stage~2 loss is defined as
\begin{equation}
\mathcal{L}_{\mathrm{Stage2}} = \frac{1}{C}\,\frac{\sum_{p} \omega(p)\,\left[y_{\mathrm{s2}}(p)-y_{\mathrm{ref}}(p)\right]^2} {\sum_{p} \omega(p) + \epsilon},
\end{equation}
where $p$ indexes voxels, $y_{\mathrm{s2}}$ denotes the density field predicted by the Stage~2 model, $y_{\mathrm{ref}}$ is the corresponding density field from the $N$-body simulation, $C$ is the number of channels, and $\epsilon$ is a small constant introduced for numerical stability.

The spatial weight function $\omega(p)$ depends on the distance between the voxels to the nearest window boundary. Specifically, for a voxel with normalized coordinates $(x,y,z)\in[0,1]^3$, the distance to the closest boundary face is defined as
\begin{equation}
d_{\mathrm{boundary}}(p) = \min\bigl(x,1-x,y,1-y,z,1-z\bigr),
\end{equation}
which vanishes for all boundary voxels, including faces, edges, and corners, and has a maximum value of $0.5$ in the center of the window. The weight then increases linearly from the boundary toward the interior and saturates to unity within a central region. The extent of this saturated region is controlled by the parameter \texttt{inner\_frac}, corresponding to $d_{\mathrm{boundary}} \ge 0.5\,\texttt{inner\_frac}$, while the minimum weight assigned to boundary voxels is set by \texttt{min\_w}.

In this work, we adopt $\texttt{inner\_frac}=0.8$, such that voxels satisfying
$d_{\mathrm{boundary}} \ge 0.5\,\texttt{inner\_frac}$ form a central region with unit weight, $\omega(p)=1$. The minimum boundary weight is set to $\texttt{min\_w}=0.2$, ensuring that all non-central voxels contribute to the loss with a reduced but non-zero weight satisfying $\omega(p)\ge 0.2$.

\subsection{Training Details}\label{sec:training}
Both stages are trained using the AdamW optimizer~\citep{Loshchilov:2017bsp}. In Stage~1, we adopt an initial learning rate of $10^{-3}$ without weight decay, and apply a StepLR scheduler with a step size of 50 epochs and a decay factor of $\gamma=0.5$. The model is trained for up to 1000 epochs, with an effective epoch size of 150 iterations. Early stopping with a patience of 20 epochs is employed based on the validation loss.

In Stage~2, the model is trained with an initial learning rate of $10^{-4}$ and a weight decay of $10^{-4}$. We adopt a cosine annealing learning rate schedule with $T_{\max}=200$ and a minimum learning rate of $10^{-7}$. Training is performed for up to 5000 epochs, with an epoch size of 500 iterations. Early stopping is applied with a patience of 100 epochs.

Stage~2 training is performed on local three-dimensional density field windows extracted from the full simulation volume using the window-based decomposition described in Section~\ref{sec:preproc}.
The windows are treated as independent training samples.
\section{Datasets and Data Processing}\label{sec:data}

In this section, we describe the numerical simulations and datasets used to train and evaluate the proposed model. We first introduce the N-body simulation suite, including simulation code, cosmological model, numerical setup, and snapshot selection. We then define the data representations and data processing procedures that specify how the simulation data are transformed and used within the proposed framework.

\subsection{N-body Simulations}\label{sec:sims}

We use \gadget\ code~\citep{Springel:2020plp} to perform the N-body simulations used for training and evaluation. The simulations are performed in a periodic cubic volume with a box size of $500~\mathrm{Mpc}/h$, containing $256^3$ dark matter particles, and with the gravitational softening length set to $4\%$ of the mean inter-particle separation. The underlying cosmological model is a flat $\Lambda$CDM model with parameter values $\Omega_{\mathrm{b}} h^2 = 0.02216$, $\Omega_{\mathrm{cdm}} h^2 = 0.1203$, $A_{\mathrm{s}} = 2.119 \times 10^{-9}$, $n_{\mathrm{s}} = 0.96$, and $h = 0.67$. Initial conditions are generated at redshift $z = 100$ using \textsc{N-GenIC} ~\citep{2015ascl.soft02003S}, adopting second-order Lagrangian perturbation theory, with the initial linear matter power spectrum computed using the Boltzmann solver \textsc{CAMB} ~\citep{Lewis:1999bs,Howlett:2012mh}.

A total of 16 snapshots are stored in logarithmically spaced redshifts between $z = 100$ and $0$, including $z = \{0, 0.2, 0.4, 0.6, 0.8, 1, 1.2, 1.5, 1.7, 2, 2.5, 3, 4, 5, 6, 100\}$. We generate 20 independent simulation realizations with different initial random seeds; 8 realizations are used for training, 2 for validation, and 10 are reserved for the final performance assessment. These snapshots form the basis for the datasets used in this work. 

\subsection{Data Processing and Reconstruction}\label{sec:preproc}

Now we describe the data processing and reconstruction steps used to transform intermediate model outputs into representations that are suitable for training and subsequent use. 

\paragraph{Density Field Window Construction}
The coarse particle positions obtained from the Stage~1 output are first converted into a three-dimensional density field, which forms the input of the Stage~2 model. We employ the cloud-in-cell (CIC) mass assignment scheme to deposit particles onto a regular Cartesian grid, yielding a discretized density field defined on an $N^3$ grid.

The resulting density field is then decomposed into overlapping spatial windows. This window-based decomposition bounds the memory cost by the window size and enables scalable processing of large simulation volumes, while remaining compatible with the voxel-wise loss function of Stage~2. The overlap between adjacent windows ensures that each spatial location is covered by multiple windows, thus reducing boundary artifacts and improving the consistency of local predictions.

The total number of windows is determined by the resolution of the grid, the size of the window, and the sliding stride. Specifically, we extract cubic density field windows of size $N_w^3$ voxels from the full $N^3$ periodic grid using a sliding-window method. Window anchors are placed every \texttt{stride} voxels along each spatial dimension, and windows extending beyond the box boundary are extracted using periodic indexing.

In practice we use $N=256$, $N_w=128$, and $\texttt{stride}=64$, which corresponds to a $50\%$ overlap in linear extent between adjacent windows.

\paragraph{Window Stitching and Reconstruction}

The Stage~2 model outputs the density-field windows at the target redshift. The final result of the full-volume density field is reconstructed using overlapping window reconstruction. For simulations with periodic initial conditions, periodic boundary conditions are applied during the reconstruction step.

For a full simulation volume discretized on an $N^3$ Cartesian grid, each predicted window with $N_w^3$ is placed back into the full volume at its corresponding spatial location and contributes to the reconstructed field through a spatial weighting function$ W(\mathbf{p})$.\footnote{Note that this weighting function is different from the one used in the Stage~2 loss.}

Let $y_i(\mathbf{p})$ denote the prediction from the $i$-th window at the local voxel coordinate $\mathbf{p}$, and let $W_i(\mathbf{p})$ be the weight of the associated window.  Then each local voxel $\mathbf{p}$ is mapped to a global voxel coordinate $\tilde{\mathbf{p}}$ in the full volume according to the spatial location of the window and the reconstructed density field is obtained as a weighted sum,
\begin{equation}
\hat{y}(\tilde{\mathbf{p}}) = \frac{\sum_i W_i(\tilde{\mathbf{p}})\, y_i(\tilde{\mathbf{p}})} {\sum_i W_i(\tilde{\mathbf{p}})},
\end{equation}
where the summation runs over all windows that cover all $\tilde{\mathbf{p}}$. 

To suppress boundary artifacts introduced by windowing, we adopt a tent-shaped weighting function $W(\mathbf{p})$, defined by
\begin{align}
W(\mathbf{p}) &= w(p_x)\, w(p_y)\, w(p_z), \\
w(p_\alpha) &= \max\!\left(0,\,1 - \frac{|p_\alpha - p_{c,\alpha}|}{(N_w-1)/2}\right),
\end{align}
where $\alpha\in\{x,y,z\}$, $\mathbf{p}=(p_x,p_y,p_z)$ denotes the voxel coordinate within a window of size $N_w^3$ and $\mathbf{p}_c=\bigl((N_w-1)/2,(N_w-1)/2,(N_w-1)/2\bigr)$ is the window center. This weighting method gives the largest weight to the interior of each window and suppresses contributions near the window boundaries, thereby reducing stitching artifacts in the reconstructed density field.

\section{Results}\label{sec:result}

We evaluate the performance of our model by comparing its predictions with the results of N-body simulations using both visual inspection and statistical measures. The results presented in this section are obtained from a model with a fixed input redshift of $z=100$ and outputs covering the redshift range $z \in [0,6]$, trained on the N-body simulation set described in Section~\ref{sec:sims}. The statistical evaluation is based on a set of probes designed to capture both Gaussian and non-Gaussian properties of the density field.

\begin{figure*}
  \centering
  \includegraphics[width=\linewidth]{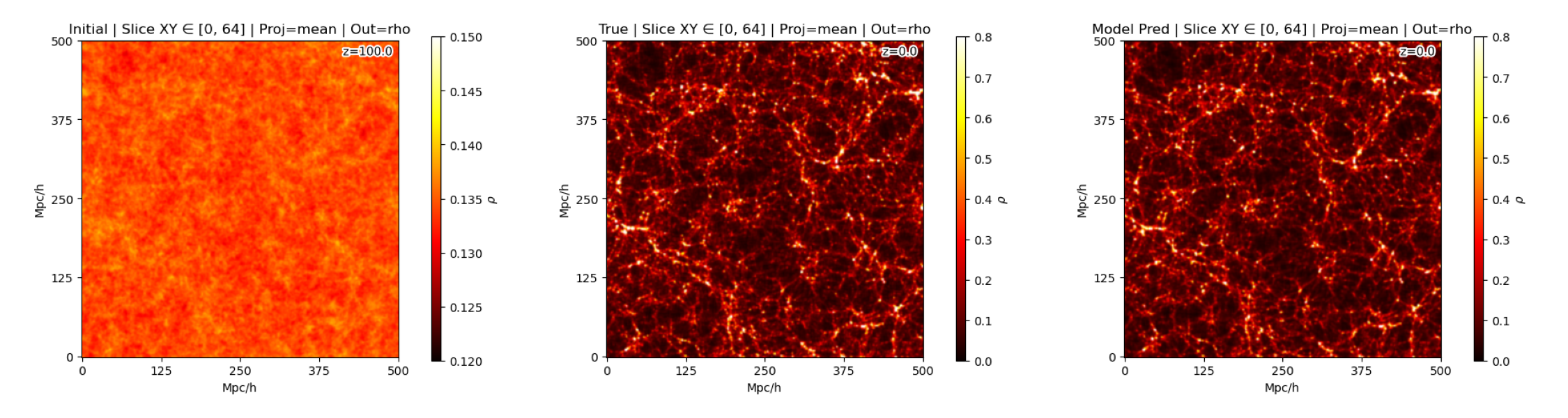}
  \caption{Two-dimensional mean density field of a slice of thickness $64\,h^{-1}\,\mathrm{Mpc}$. The left panel shows the input initial condition at $z=100$. The middle panel shows the reference N-body result at $z=0$, and the right panel shows the corresponding model prediction.}
  \label{fig:visual_comparison}
\end{figure*}

\subsection{Visualization Comparison}\label{sec:vision}

We begin with a visual comparison between the density fields predicted by the model and the reference N-body results. Figure~\ref{fig:visual_comparison} shows an example of a two-dimensional mean density field of the three-dimensional density field with a thickness of $64\,h^{-1}\,\mathrm{Mpc}$. The predicted field closely matches the N-body result, reproducing both large-scale structures and small-scale features, with no obvious discrepancies visible by eye.

\subsection{One-point Statistics}\label{sec:1pt}

One-point statistics probe the local distribution of density values and are sensitive to nonlinear and non-Gaussian features arising from gravitational evolution. We evaluate the one-point probability distribution function (PDF) of the over-density field $\delta(\mathbf{x})$, defined as
\begin{equation}
\delta(\mathbf{x}) = \frac{\rho(\mathbf{x})}{\bar{\rho}} - 1 ,
\end{equation}
where $\rho(\mathbf{x})$ is the density field of matter and $\bar{\rho}$ is its mean value.

\begin{figure}
  \centering
  \includegraphics[width=\linewidth]{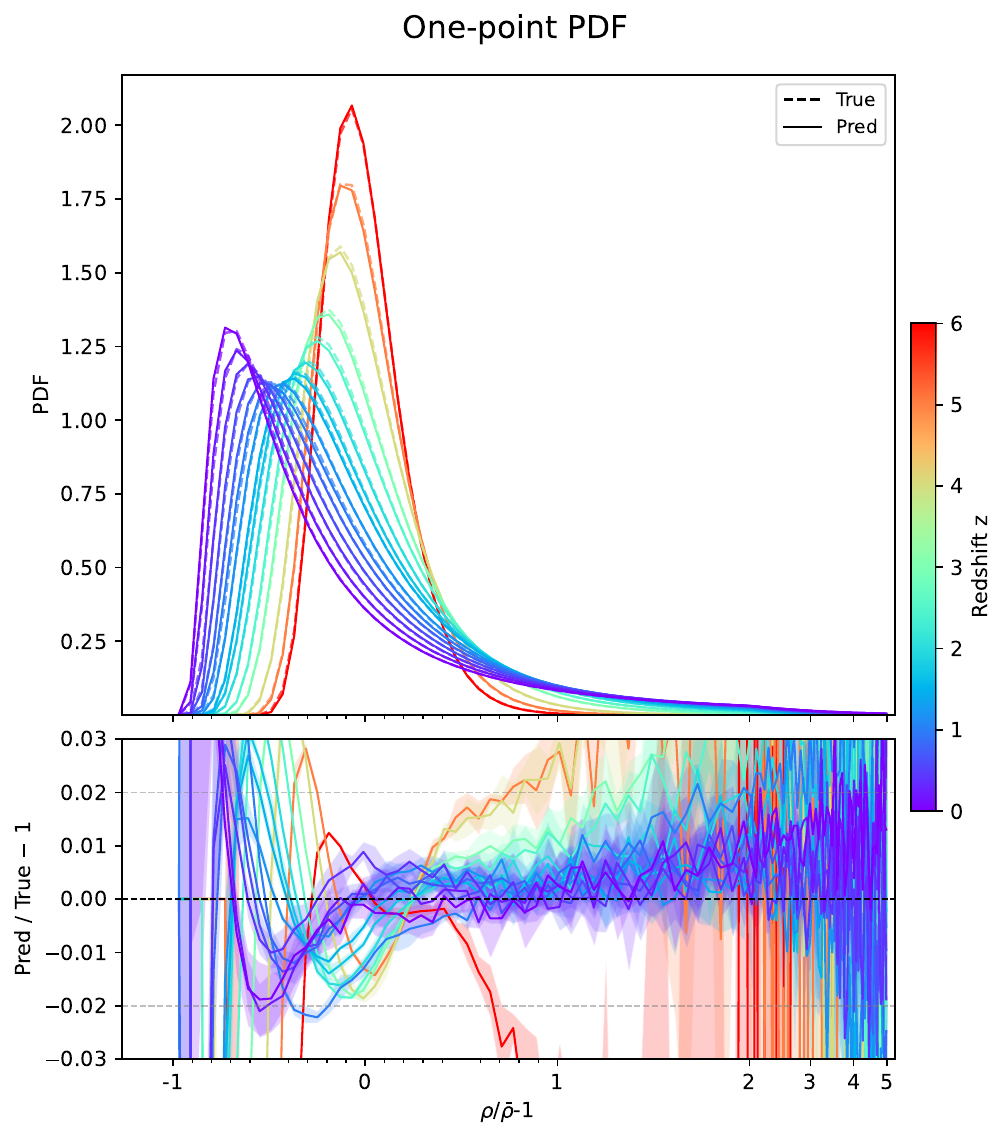}
  \caption{One-point probability distribution functions (PDFs) of the density field, smoothed with a Gaussian kernel of radius $2\,h^{-1}\,\mathrm{Mpc}$, measured from the model predictions and the reference N-body simulations. The PDFs are averaged over 10 independent test simulations. Colors represent different redshifts, as indicated by the color bar on the right. Upper panel: mean PDFs, where dashed lines represent the reference N-body results and solid lines denote the model predictions. Lower panel: relative errors with a $1\sigma$ error band. The $\pm 2\%$ reference levels are indicated by gray dashed lines, and the zero level by a black dashed line.}
  \label{fig:pdf_results}
\end{figure}

Figure~\ref{fig:pdf_results} shows the PDF of the model predictions compared with the N-body reference results after Gaussian smoothing on a scale of $2\,h^{-1}\,\mathrm{Mpc}$. Overall, the predicted PDFs agree well with the reference results. The relative error remains below the $2\%$ level over the intermediate density range, roughly $\delta \in [-0.5,\,2]$, and becomes larger toward both lower- and higher-density regimes. This trend is less pronounced at higher redshifts, where the PDF is more symmetric and closer to a Gaussian distribution due to the weaker nonlinear evolution.

The density PDF captures deviations from Gaussianity induced by nonlinear gravitational evolution, including the development of high-density peaks associated with collapsed structures and low-density regions corresponding to cosmic voids. At higher redshifts, the PDF remains closer to Gaussian and more symmetric, reflecting the weaker level of nonlinear evolution. The overall agreement between the predicted and reference PDFs across redshifts indicates that the model successfully reproduces the one-point statistics of the matter distribution, with deviations at the few-percent level over the intermediate density range.

\subsection{Two-point Statistics}\label{sec:2pt}
Two-point statistics, including the two-point correlation function and its Fourier-space counterpart, the power spectrum, are among the most widely used statistical measures in cosmology. They are primarily sensitive to the Gaussian features of the large-scale structure and provide a quantitative description of clustering across different spatial scales. In this work, we assess the model performance using the three-dimensional matter power spectrum constructed from the over-density field $\delta(\mathbf{x})$.

The matter power spectrum is defined as
\begin{equation}
P(k) = \langle |\delta(\mathbf{k})|^2 \rangle ,
\end{equation}
where $\delta(\mathbf{k})$ is the Fourier transform of $\delta(\mathbf{x})$, and $k$ denotes the wavenumber of density fluctuations. To quantify the amplitude accuracy of the model predictions, we introduce the transfer function,
\begin{equation}
T(k) = \sqrt{\frac{P_{\mathrm{pred}}(k)}{P_{\mathrm{ref}}(k)}} ,
\end{equation}
where $P_{\mathrm{pred}}(k)$ and $P_{\mathrm{ref}}(k)$ denote the power spectra of the model predictions and the reference N-body simulations, respectively.

\begin{figure}
  \centering
  \includegraphics[width=\linewidth]{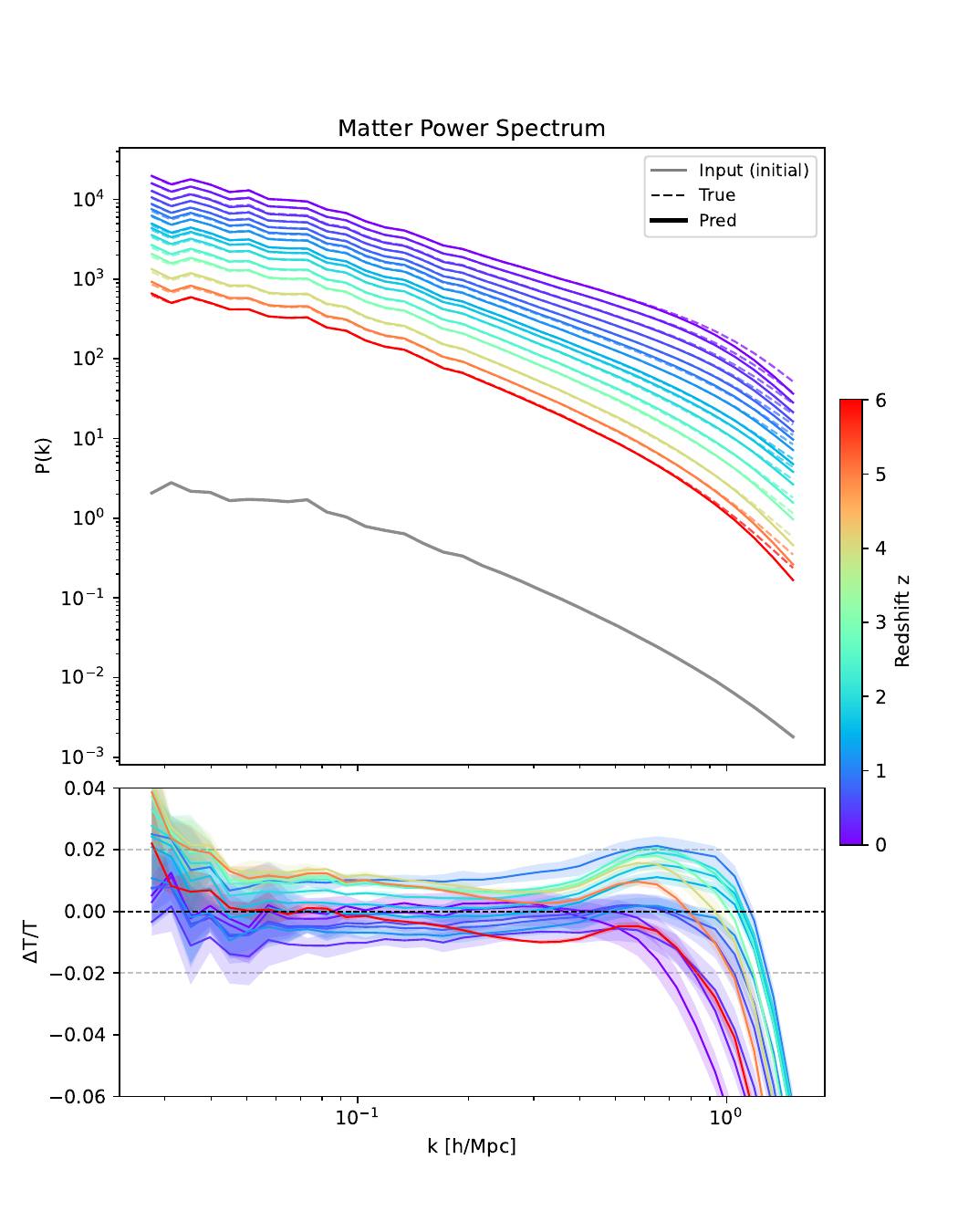}
  \caption{Matter power spectra measured from the density fields predicted by the model and from those generated by the reference N-body simulations, averaged over 10 independent test simulations. Colors represent different redshifts, as indicated by the color bar on the right. Upper panel: mean matter power spectra, where dashed lines represent the reference N-body results and solid lines denote the model predictions. The gray solid line shows the power spectrum of the initial conditions. Lower panel: relative transfer function errors with a $1\sigma$ error band. The $\pm 2\%$ reference levels are indicated by gray dashed lines, and the zero level by a black dashed line.}
  \label{fig:pk_results}
\end{figure}

As shown in Fig.~\ref{fig:pk_results}, the relative transfer function errors remain below $3\%$ up to $k \simeq 0.7\,h\,\mathrm{Mpc}^{-1}$. At smaller scales, the agreement gradually degrades as nonlinear effects become more significant. Predictions at higher redshifts show better performance on small scales, consistent with lower nonlinear evolution at earlier cosmic times.

\begin{figure}
  \centering
   \includegraphics[width=\linewidth]{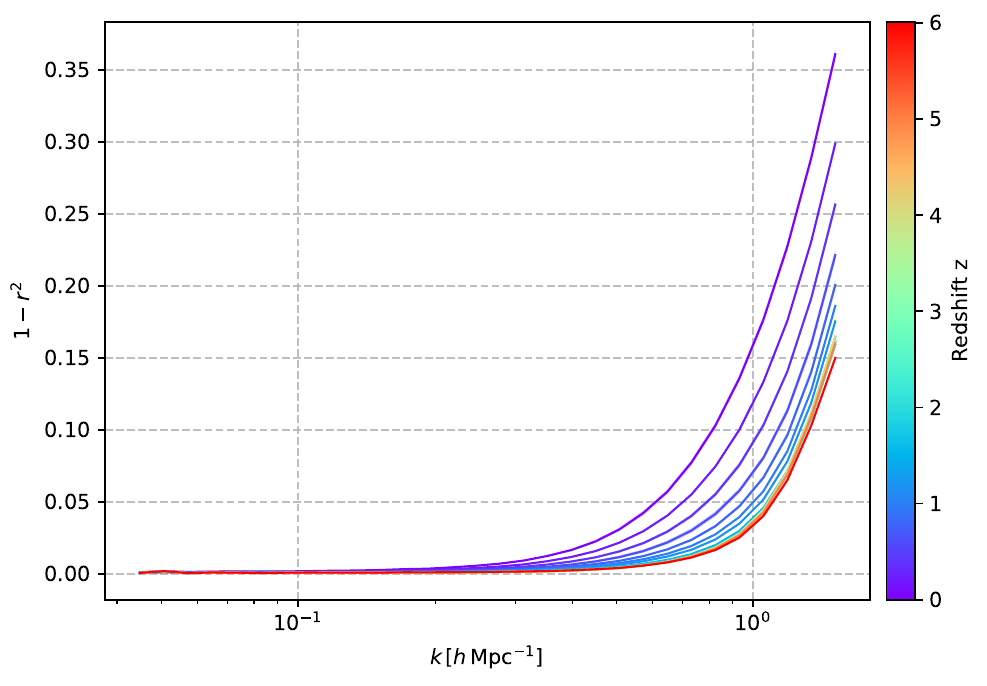}
  \caption{The stochasticity $1 - r^2$ between the density fields predicted by the model and the reference N-body simulations, averaged over 10 test simulations. Colors indicate different redshifts, as shown by the color bar on the right.}
  \label{fig:crosspk_results}
\end{figure}

We further quantify phase accuracy using the cross-correlation coefficient,
\begin{equation}
r(k) =
\frac{P_{\mathrm{cross}}(k)}
{\sqrt{P_{\mathrm{pred}}(k)\,P_{\mathrm{ref}}(k)}} ,
\end{equation}
where $P_{\mathrm{cross}}(k)$ denotes the cross power spectrum between the predicted and reference density fields. The coefficient $r(k)$ measures the phase alignment of Fourier modes between the two fields. In practice, we present $1 - r^2(k)$, often referred to as stochasticity, which provides a convenient measure of phase decorrelation and vanishes in the case of perfect phase agreement. The results are shown in Fig.~\ref{fig:crosspk_results}. At $k \simeq 1\,h\,\mathrm{Mpc}^{-1}$, models evaluated at redshifts $z \gtrsim 1$ exhibit phase errors below the $5\%$ level, while at lower redshifts the deviations remain within approximately $15\%$.

\subsection{Peak Statistics}\label{sec:peak}

Peak statistics quantify the abundance of local maxima in the density field as a function of their significance and provide sensitivity to higher-order non-Gaussian features induced by nonlinear gravitational evolution. We identify peaks in the three-dimensional overdensity field after smoothing with a Gaussian filter of scale $R$. A peak is defined as a grid cell whose overdensity exceeds that of all its neighboring cells.

The peak height is expressed in units of the local standard deviation, denoted as the signal-to-noise ratio,
\begin{equation}
\mathrm{S/N} = \frac{\delta}{\sigma(R)},
\end{equation}
where $\sigma(R)$ is the standard deviation of the density field smoothed on scale $R$. We consider smoothing scales of $R = 2,\,4,$ and $8\,h^{-1}\,\mathrm{Mpc}$. The peak abundance $n_{\mathrm{pk}}$ is then evaluated as a function of $\mathrm{S/N}$ for each smoothing scale, probing nonlinear clustering and high-density regions that are not fully captured by two-point statistics.

\begin{figure} 
\centering 
\includegraphics[width=\linewidth]{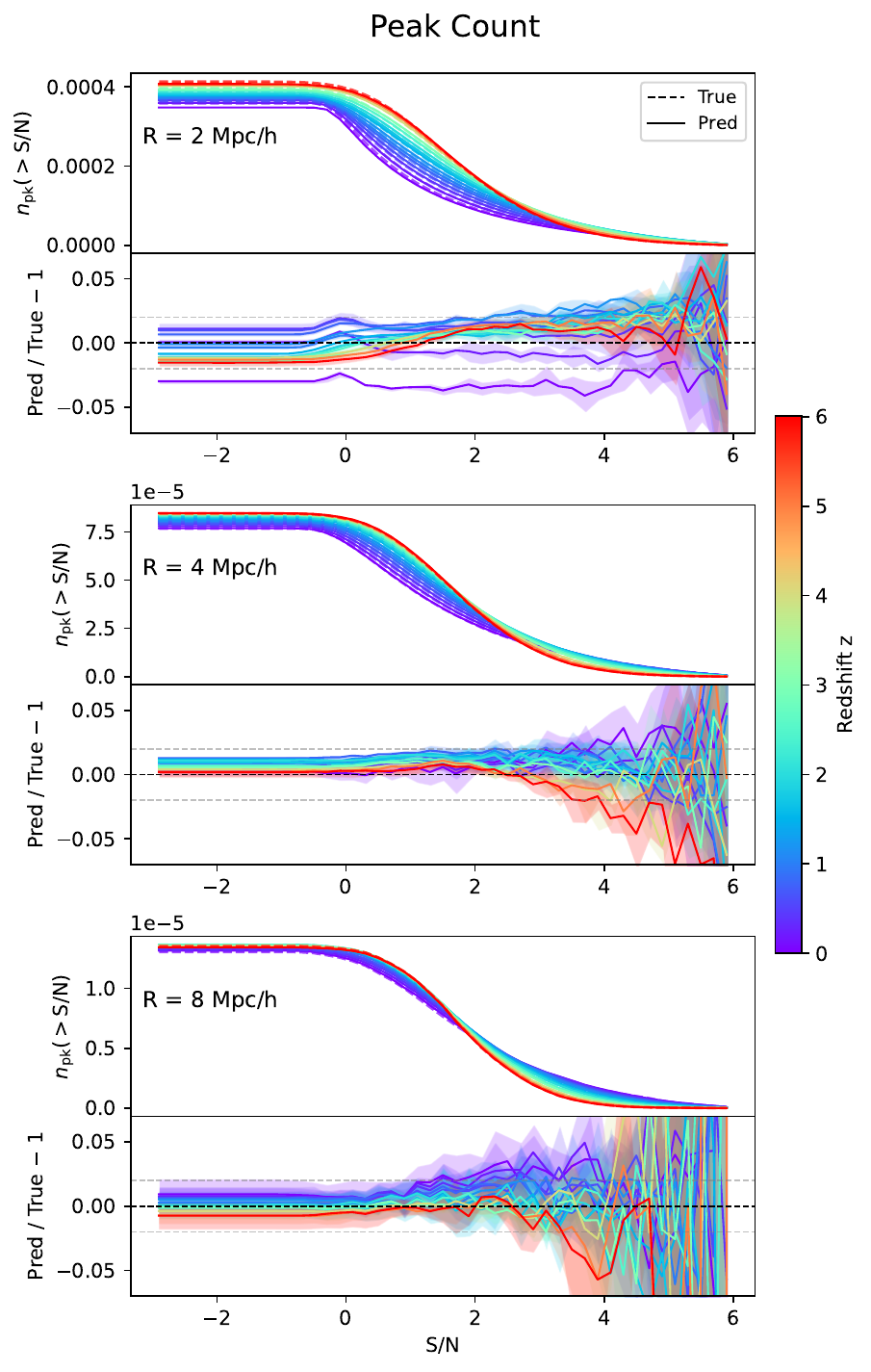} 
\caption{Peak abundance as a function of $\mathrm{S/N}$ for different smoothing scales. The upper panel of each subplot shows the absolute peak abundance $n_{\mathrm{pk}}$, averaged 10 test simulations, with solid lines denoting the model predictions and dashed lines indicating the reference N-body results. The lower panel shows the corresponding relative errors, with a $1\sigma$ error band and $\pm 2\%$ reference levels indicated by gray dashed lines. Each subplot corresponds to a different smoothing scale, $R = 2,\,4,$ and $8\,h^{-1}\,\mathrm{Mpc}$. The redshift color mapping is indicated by the color bar on the right.} 
\label{fig:peak_results} 
\end{figure}

The results are shown in Fig.~\ref{fig:peak_results}. For all smoothing scales, the relative error in the peak abundance remains below the $10\%$ level for $\mathrm{S/N} \lesssim 3$. At higher peak significance, the discrepancies increase, reflecting the enhanced impact of nonlinear evolution and the reduced statistical weight of rare high-density peaks. Overall, the model reproduces the peak statistics of the density field with good accuracy across redshifts.

\begin{figure*}
  \centering
  \includegraphics[width=\linewidth]{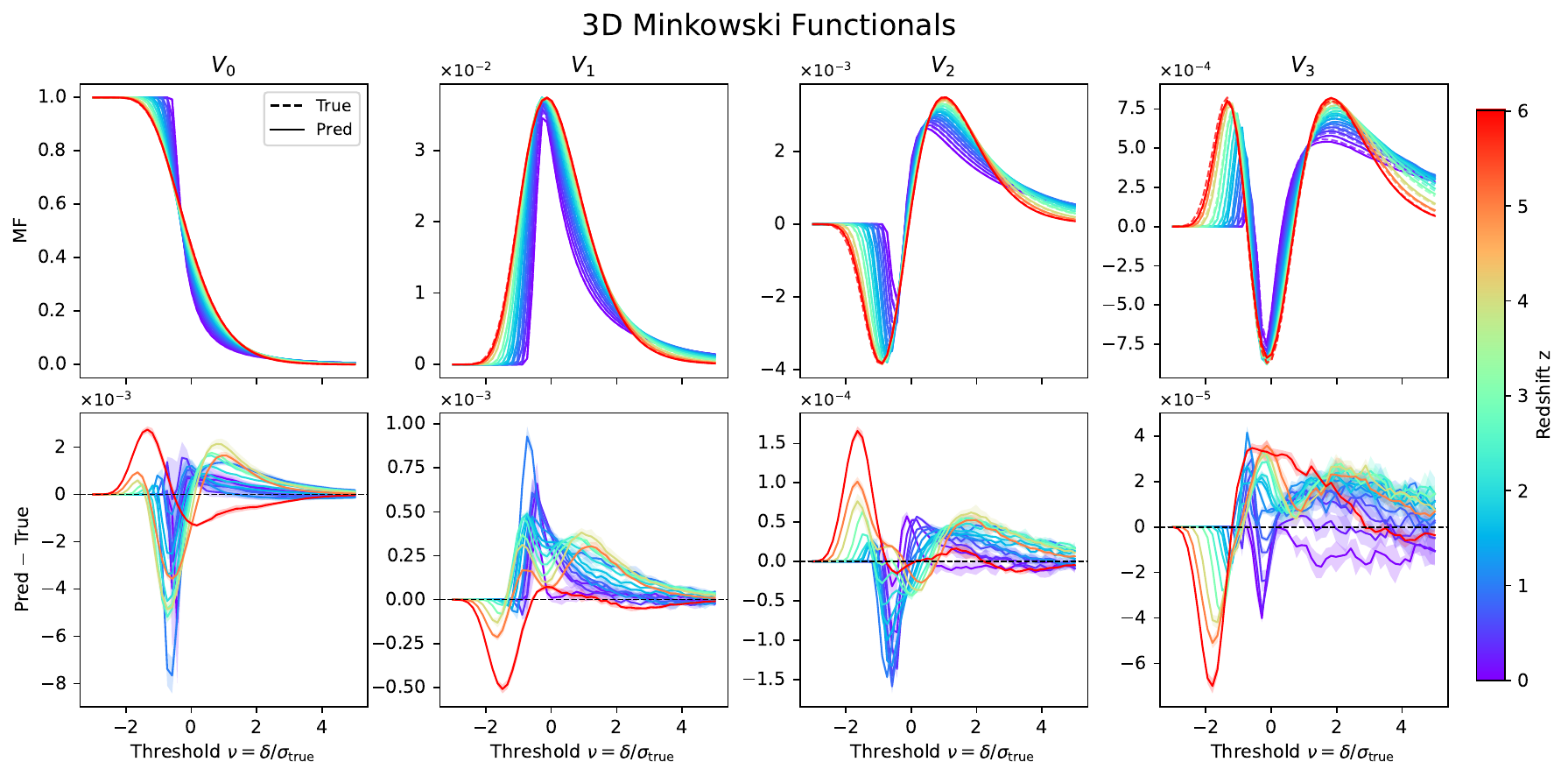}
  \caption{Three-dimensional Minkowski functionals. For each functional ($V_0$, $V_1$, $V_2$, and $V_3$), the upper panels show the absolute values as functions of the density threshold, averaged over 10 test simulations, with solid lines denoting the model predictions and dashed lines indicating the reference N-body results. The lower panels show the differences between the predictions and the reference results, together with a $1\sigma$ error band. The redshift dependence is encoded by the color bar on the right. All density fields are smoothed with a Gaussian filter of scale $2\,h^{-1}\,\mathrm{Mpc}$.}
  \label{fig:mf_results}
\end{figure*}

\subsection{Three-dimensional Minkowski Functionals}\label{sec:mf}

Three-dimensional Minkowski functionals provide a set of morphological descriptors that characterize the global geometry and topology of the density field beyond local density values. In this work, the Minkowski functionals are measured numerically following the integral-geometry approach developed by \citet{1997MNRAS.284...73K} and \citet{Gleser:2006su}, which allows a direct evaluation of Minkowski functionals on a discretized three-dimensional lattice. The four Minkowski functionals considered here are the volume fraction ($V_0$), the surface area ($V_1$), the integrated mean curvature($V_2$) and the Euler characteristic ($V_3$).

In practice, Minkowski functionals are evaluated on excursion sets of the density field. For a given threshold $\nu$, the excursion set is defined as the collection of all spatial regions where the smoothed density field higher than threshold. As the threshold is varied, the excursion set undergoes a continuous morphological evolution. The Minkowski functionals quantify this evolution by measuring how the volume, surface geometry, curvature, and topology of the excursion set change as a function of $\nu$.

The numerical measurement method is based on expressing the local geometric properties of isodensity surfaces in terms of the principal curvatures, which are constructed from first and second derivatives of the density field. Using the formalism introduced by \citet{Koenderink1984}, the local curvatures are written in terms of geometric invariants, commonly referred to as the Koenderink invariants. For a three-dimensional scalar field $u(\mathbf{x})$, the sum and product of the principal curvatures $\kappa_1$ and $\kappa_2$ are given by
\begin{align}
\kappa_1 + \kappa_2 &=
\frac{\epsilon_{ijm}\epsilon_{klm}\,u_{,i}\,u_{,jk}\,u_{,l}}
{\left(u_{,n}u_{,n}\right)^{3/2}} , \label{eq:kappa_sum} \\[6pt]
\kappa_1 \kappa_2 &=
\frac{\epsilon_{ijk}\epsilon_{lmn}\,u_{,i}\,u_{,l}\,u_{,jm}\,u_{,kn}}
{2\left(u_{,p}u_{,p}\right)^{2}} , \label{eq:kappa_prod}
\end{align}
where $u_{,i}$ and $u_{,ij}$ denote the first and second spatial derivatives of the field, and $\epsilon_{ijk}$ is the Levi-Civita tensor.

With these local curvature measures, the surface integrals defining the Minkowski functionals can be replaced by volume integrals over the entire domain. The three non-trivial Minkowski functionals in three dimensions are then given by
\begin{align}
V_1(\nu) &= \frac{1}{6V}
\int_V \mathrm{d}^3x\,
\delta_\mathrm{D}\!\left(\nu\sigma - u(\mathbf{x})\right)
\left|\nabla u(\mathbf{x})\right| , \label{eq:mf_v1} \\[6pt]
V_2(\nu) &= \frac{1}{6\pi V}
\int_V \mathrm{d}^3x\,
\delta_\mathrm{D}\!\left(\nu\sigma - u(\mathbf{x})\right)
\left|\nabla u(\mathbf{x})\right|
\left[\kappa_1(\mathbf{x}) + \kappa_2(\mathbf{x})\right] , \label{eq:mf_v2} \\[6pt]
V_3(\nu) &= \frac{1}{4\pi V}
\int_V \mathrm{d}^3x\,
\delta_\mathrm{D}\!\left(\nu\sigma - u(\mathbf{x})\right)
\left|\nabla u(\mathbf{x})\right|
\kappa_1(\mathbf{x})\kappa_2(\mathbf{x}) , \label{eq:mf_v3}
\end{align}
where $V$ is the total volume of the domain, $\nu$ is the density threshold in units of the standard deviation $\sigma$ of the field, and $\delta_\mathrm{D}$ denotes the Dirac delta function.

This numerical formulation enables a robust and assumption-free measurement of Minkowski functionals directly from discretized density fields, making it well suited for quantifying the non-Gaussian and nonlinear morphological features of large-scale structure. The results are shown in Fig.~\ref{fig:mf_results}, where the density fields are smoothed with a Gaussian filter of scale $2,h^{-1},\mathrm{Mpc}$.

Overall, the model predictions are in good agreement with the reference N-body results, with discrepancies typically more than one order of magnitude smaller than the absolute values of the Minkowski functionals. The largest deviations occur around thresholds $\nu \in [-2,1]$. 
As shown in the lower panel of Fig.~\ref{fig:mf_results}, the largest deviation from the true value occurs at high redshifts around \(\nu \approx -2\), corresponding to the peaks and troughs of the reddish curves. As redshift decreases, this feature shifts toward higher thresholds, with the bluish curves exhibiting peaks and troughs near \(\nu \approx 0\).

Regarding the overall shape, all four Minkowski functionals exhibit a systematic shift toward the high-density tail as the redshift decreases. This trend reflects the increasing dominance of nonlinear clustering in shaping the topological properties of the density field at late times.

\subsection{Runtime and Memory Performance}\label{sec:runtime}

We evaluated the runtime and memory performance of the proposed pipeline on a single NVIDIA RTX 4090 GPU. All GPU-based inference is performed with a maximum memory usage of approximately $4\,\mathrm{GB}$. Data processing steps, including CIC mass assignment and sliding-window construction, are executed on the CPU.

For a full-volume inference at a single redshift, starting from an initial particle distribution with $256^3$ particles, the trained model produces the corresponding $256^3$ density grid output in approximately $25\,\mathrm{s}$. This runtime includes both Stage~1 particle-based inference and Stage~2 density-field refinement, as well as the associated data transfers between the CPU and GPU. Performance results
are summarized in Table.~\ref{tab:runtime}.

The proposed framework is inherently scalable. By processing particles in batches during Stage~1 and assembling overlapping density windows during Stage~2, the model can be applied to larger simulation volumes and to domains with irregular geometries without increasing GPU memory requirements. This design enables efficient inference on volumes beyond the size used during training, making the method suitable for large-scale applications where memory constraints are critical.

\begin{table}
\centering
\caption{Runtime and memory performance for a single full-volume inference from a $256^3$ particle input to a $256^3$ density grid at one target redshift.}
\label{tab:runtime}
\begin{tabular}{l|l}
\hline
\hline
GPU & NVIDIA RTX 4090 \\
Peak GPU memory & $\sim 4\,\mathrm{GB}$ \\
\hline
Input & $256^3$ particles \\
Output & $256^3$ density grid\\
\hline
End-to-end inference time & $\sim 25\,\mathrm{s}$ \\
\hline
\hline
\end{tabular}
\end{table}

\section{Conclusion}\label{sec:conclusion}

In this work, we have presented a neural network framework for accelerating N-body simulations by predicting the nonlinear evolution of the matter density field from an initial particle distribution. The proposed approach adopts a two-stage design, in which particle-based inference is first performed to capture large-scale gravitational evolution, followed by a density-field refinement stage that restores small-scale nonlinear structures.

We evaluate the model using a comprehensive set of statistical diagnostics, including two-point statistics, one-point probability distributions, peak statistics, and three-dimensional Minkowski functionals. Across these probes, the predicted density fields show good agreement with reference N-body simulations over a wide range of redshifts. In particular, the model reproduces not only Gaussian clustering statistics but also higher-order non-Gaussian and topological features induced by nonlinear gravitational evolution.

Our model demonstrates high efficiency with a relatively small number of training simulations and modest computational requirements. In our experiments, only 8 simulation realizations are used for training, and the model can be executed on a single consumer-grade GPU (NVIDIA RTX 4090) with approximately $4\mathrm{GB}$ of memory usage. Besides the low memory requirement, the inference time is fast: starting from an initial $256^3$ particle distribution at $z=100$, the model generates a full $256^3$ density grid at the target redshift in approximately 25~seconds on a single GPU. 

Furthermore, by processing particles in batches and assembling overlapping density windows, the method can be naturally extended to large simulation volumes and irregular geometries without increasing GPU memory requirements.

An important advantage of the proposed framework is that it is not intrinsically tied to a specific high-redshift initial condition. By operating directly on particle distributions, the model can, in principle, be initialized at intermediate redshifts, thereby reducing sensitivity to poorly constrained early-time physics. This flexibility opens up new possibilities for applications in which only late-time or observationally constrained density fields are available, such as survey-based reconstructions and light-cone simulations.

The combination of flexibility, accuracy, and computational efficiency makes the proposed model a promising tool for applications that require large ensembles of cosmological realizations or rapid forward modeling, including weak gravitational lensing analyses, non-Gaussian statistics, and cosmological parameter inference.

In future work, we will train the model on a larger set of N-body simulations and higher-resolution data, as the current study uses only eight simulations with $256^3$ particles primarily to demonstrate the capability of the proposed framework. The present prediction accuracy is therefore likely limited by both the size and the resolution of the training dataset. Increasing the number of training simulations would improve the sampling of cosmic variance and the diversity of nonlinear structures, while higher-resolution simulations could provide more accurate small-scale information and potentially improve the prediction accuracy at higher $k$. Further gains may also be achieved by exploring alternative network architectures and training strategies. We also plan to extend the framework to modified gravity simulations, multi-component matter fields, and applications such as light-cone generation and observational data analysis.


\section*{Acknowledgments}
We thank Baojiu Li and Yin Li for the insightful discussion. 
This work is supported by the National Natural Science Foundation of China Grants No. 12541301. 
We acknowledge the use of the Katana high-performance computational cluster at UNSW Sydney~\citep{KatanaHPC} for carrying out the early-stage N-body simulations used in this work. We also acknowledge the use of ChatGPT for language polishing and assistance with code organization.

\bibliography{refs}
\bibliographystyle{aasjournalv7}

\end{document}